\documentclass[twocolumn,showpacs,preprintnumbers,superscriptaddress,amsmath,amssymb]{revtex4-1}

\usepackage{epsfig}
\usepackage{graphicx}% Include figure files
\usepackage{dcolumn}% Align table columns on decimal point
\usepackage{bm}% bold math
\usepackage{times}
\usepackage{xcolor}

\begin{document}

%\preprint{APS/123-QED}

\title{Dynamics of social contagions with limited contact capacity}

\author{Wei Wang}
\affiliation{Web Sciences Center, University of Electronic
Science and Technology of China, Chengdu 610054, China}

\author{Panpan Shu}
\affiliation{Web Sciences Center, University of Electronic
Science and Technology of China, Chengdu 610054, China}

\author{Yu-Xiao Zhu}
\affiliation{Web Sciences Center, University of Electronic
Science and Technology of China, Chengdu 610054, China}

\author{Ming Tang} \email{tangminghuang521@hotmail.com}
\affiliation{Web Sciences Center, University of Electronic
Science and Technology of China, Chengdu 610054, China}

\author{Yi-Cheng Zhang}
\affiliation{Department of Physics, University of Fribourg,
Chemin du Mus\'{e}e 3, 1700 Fribourg, Switzerland}

\date{\today}

\begin{abstract}
Individuals are always limited by some inelastic resources, such as time and energy,
which restrict them to dedicate to social interaction and limit their contact
capacity. Contact capacity plays an important role in dynamics of social
contagions, which so far has eluded theoretical analysis.
In this paper, we first propose a non-Markovian model to understand
the effects of contact capacity on  social contagions, in which each individual
can only contact and transmit the information to a finite number of neighbors.
We then develop a
heterogeneous edge-based compartmental theory for this model,
and a remarkable agreement with simulations is obtained. Through theory and simulations,
we find that enlarging the contact capacity makes the network more fragile to behavior
spreading. Interestingly, we find that both the
continuous and discontinuous dependence of the final adoption size on the
information transmission probability can arise. And there is a
crossover phenomenon between the two types of dependence.
More specifically, the crossover phenomenon can be
induced by enlarging the contact capacity only when the degree exponent is above
a critical degree exponent, while the the final behavior adoption size always grows
continuously for any contact capacity when degree exponent is below the critical
degree exponent.

\end{abstract}

\maketitle

\textbf{Recent empirical studies of Facebook communication networks, scientific cooperation
networks and sexual contact networks suggest that individuals' activities are limited
by the time, funds, energy and other inelastic resources. Thus, individuals exhibit limited
contact capacity (i.e., individuals can only communicate
or interact with a finite number of neighbors during a short time)
in the dynamics of epidemic and behavior. Previous studies have proven
that limited contact capacity enlarges the epidemic outbreak threshold and makes the theoretical prediction deviate from simulation
results more easily. Unfortunately, a systematical investigation the
effects of contact capacity on the dynamics of social contagions is still
lacking. To fill this gap, we first propose a non-Markovian behavior spreading model
with limited contact capacity. Then, we develop a novel
heterogeneous edge-based compartmental theory for this proposed model, and verify the
effectiveness of this suggested theory based on large number of simulations.
Through theory and simulations, we find that the final behavior adoption size increases
with the contact capacity. Strikingly, we uncover a crossover phenomenon in which the
dependence of the final adoption size on the information transmission probability can
change from being continuous to being discontinuous. We
find a critical degree exponent above which the crossover phenomenon can be induced by
enlarging the contact capacity. However, the final adoption size always
grows continuously for any contact capacity when degree exponent is below this
critical degree exponent. Our
results help us to have a deeper understanding of the effects of contact capacity
on social contagions, and the developed theory could be applied to other analogous dynamical
processes (e.g., information diffusion and cascading failure).}

\section{Introduction} \label{sec:intro}
Humans are the basic constituents of the society, and every individual can interact with
his/her family, friend and peers. These interactions among individuals can induce some
interesting collective behavior, such as, spontaneous formation of a common language or
culture, emergence of consensus about a specific issue, and the adoptions of
innovation, healthy or microfinance behavior. Understanding the mechanisms or regularities
behind these collective behavior has led to a booming subfield of research
in complex network science -- social contagions, which has attracted much attention in
recent years~\cite{Castellano2009,Barrat2008,Vicsek2012}.

Statistical physics approaches were widely used to investigate social contagions.
On the one hand, scientists used these methods to analyse large
databases of social contagions, and revealed that reinforcement effect
widely exists~\cite{Pastor-Satorras2014}. The
reinforcement effect means that individual adopting a behavior is based on the
memory of the cumulative behavioral information
that he/she received from his/her neighbors.
Centola established the artificially
structured online communities to study health behavior spreading, and found that
the reinforcement effect significantly increases the adoption of a new health
behavior~\cite{Centol2010,Centol2011}. The reinforcement effect also exists in the
adoptions of Facebook~\cite{Ugander2012} and Skype~\cite{Karsai2014a} services. On the other
hand, researchers proposed some novel models with reinforcement effect to describe
the dynamics of social contagions. Among these models, linear threshold
model~\cite{Watts2002,Glesson2007,Brummitt2012} is a famous one, and it is a deterministic
model (i.e., a trivial case of Markovian process) once the network topology and initial seeds are fixed. In this model, an individual will adopt the behavior once the current
fraction of his/her adopted neighbours is larger than a static threshold. The linear
threshold model induces that the final behavior adoption size first grows continuously
and then decreases
discontinuously with the increasing of mean degree for vanishing small fraction of
seeds. Another more realistic way to incorporate the reinforcement effect is whether an
individual adopts the behavior should take his/her cumulative pieces of behavioral
information into consideration~\cite{Dodds2004,Dodds2005,Wang2015a,Zheng2013,Chung2014}.
In this case, the dynamics is a non-Markovian process,
which makes it more difficult to develop an accurate theory. Wang \emph{et al}
proposed a non-Markovian
behavior spreading model, and found that the dependence of final behavior
adoption size on information transmission probability can change from being discontinuous
to being continuous under dynamical or structural parameters perturbation~\cite{Wang2015a}.

Recently, scholars found that individuals exhibit limited contact capacity
(i.e., individuals can only communicate or interact with a finite
number of neighbors during a short time)
since the inelastic resources (e.g., time, funds, and energy) restrict them to dedicate to
social interaction from empirical analysis~\cite{Haerter2012,Miritello2013,Holme2012}.
In Facebook communication networks, Golder \emph{et al} revealed that users only communicate
with a small number of people even though they have many declared friends~\cite{Golder2007}.
In scientific cooperation networks, a scientist exchanges knowledge with only a fraction of
his/her cooperators in a paper~\cite{Perra2012,Karsai2014}. In sexual contact networks,
individuals can not have sexual intercourse with his/her all sexual partners in
a very short time due to the limitation of morality and
physiology~\cite{Liljeros2001,Liljeros2003}.
Researchers have studied the effects of contact capacity on some
Markovian dynamics (i.e., epidemic spreading)~\cite{Cui2014,Yang2007,Castellano2006}.
They found that the epidemic outbreak threshold increases when the contact capacity is
limited~\cite{Yang2007}. Meanwhile, each connection (edge) has distinct effective spreading
probability (to be defined in Sec.~\ref{sec:theory}), which
makes the theoretical prediction deviate from simulation results more easily, especially in
the case of strong structural heterogeneity.

For the dynamics of social contagions, whether an
individual adopts a behavior behavior or not is determined by the cumulative pieces
of behavioral information that he/she has received from neighbors~\cite{Centol2011,Wang2015a}.
Once the contact capacity is limited, the behavioral information transmission
will be limited, thus further affects the dynamics of social contagions. However,
a systematic study to understand the effects of contact capacity on
dynamics of social contagions is still lacking. In this paper, we try to address
how the contact capacity affects the behavior spreading dynamics.
We first propose a
non-Markovian behavior spreading model with limited contact capacity, in which each
adopted individual tries to transmit the behavioral information to a finite number of his/her
neighbors. In order to understand, quantitatively, the effects of contact
capacity on behavior spreading, we develop a heterogeneous edge-based compartmental
theory. We find that the final
behavior adoption size increases with the contact capacity. More interestingly, the
crossover phenomenon in which the dependence of the final adoption size on
the information transmission probability can change from being continuous to being
discontinuous. By enlarging the contact capacity,
the crossover phenomenon can be induced only when the degree exponent is above a critical
critical degree exponent. However, the final
adoption size always grows continuously for any contact capacity when degree
exponent is below the critical degree exponent. The theoretical
results from the suggested method can accurately predict the above results.

The paper is organized as follows. In Sec.~\ref{sec:model}, we describe the behavior
spreading model with limited contact capacity. We develop the heterogeneous edge-based
compartmental theory in Sec.~\ref{sec:theory}. In Sec.~\ref{sec:sim}, we verify the effectiveness
of the theory through large number of simulations. Finally, we present
conclusions and discussions in Sec.~\ref{sec:dis}.

\section{Behavior Spreading Model} \label{sec:model}
We consider the behavior spreading on uncorrelated
configuration networks~\cite{Newman2010,Catanzaro2005} with $N$ individuals (nodes) and degree
distribution $P(k)$.
We use a generalized model SAR (susceptible-adopted-recovered)
model~\cite{Wang2015a} to describe behavior spreading on networks. At each time step,
each individual can be in one of the three different states: susceptible, adopted,
or recovered. In the susceptible state, an individual does not
adopt the behavior. In the adopted state, an individual
adopts the behavior and tries to transmit the information
to his/her selected neighbors. In the recovered state, an individual
loses interest in the behavior and will not transmit the information
further. Each individual holds a static adoption threshold $\kappa$,
which reflects the criterion (wills) of an individual to adopt the behavior.

Initially, a fraction of $\rho_0$ individuals (nodes) are randomly selected to be
in the adopted state (seeds),
while other individuals are in the susceptible states. All susceptible individuals do not
know any information about this behavior, in other words, the cumulative pieces of
information is zero initially for all susceptible individuals. At each time step, each adopted
individual $v$ with $k^\prime$ neighbors randomly chooses $f(k^\prime)$ number of neighbors
due to the contact capacity is limited,
and tries to transmit the information to each selected neighbors with probability $\lambda$.
Note that the function $f(k^\prime)$ represents the contact capacity
of $v$, the larger value of $f(k^\prime)$, the more neighbors can receive the information
from him/her. If $f(k^\prime)<k^\prime$, the contact capacity of individual $v$ is limited.
Once the contact capacity of $v$ is larger than his/her degree, we let he/she transmit
information to his/her all neighbors. If $v$ transmits the information to $u$ successfully,
the cumulative pieces of information $m$ that $u$ ever received will increase by $1$, and
the information can not be transmitted between $u$ and $v$ in the following spreading process
(i.e., redundant information transmission on the edge is forbidden). If $m$ is larger
than the adoption threshold $\kappa$, individual $u$ becomes adopted in the next
time step. From the mentioned procedures of susceptible individuals becoming adopted,
we learn that the dynamics of social contagion is a non-Markovian process.
The adopted individuals then lose interest in the behavior and enters into recovered
with probability $\gamma$. Individuals in the recovered state do not take part
in the spreading process. The dynamics terminates once all adopted individuals become
recovered.

\section{Heterogeneous Edge-Based Compartmental Theory} \label{sec:theory}
The non-Markovian behavior spreading model with limited contact capacity
described in Sec.~\ref{sec:model} makes theoretical
prediction from the classical theory (e.g., heterogeneous mean-field theory) deviate from
simulation results easily. On the one hand, in this
proposed model, whether a susceptible individual adopts the behavior or not is dependent
on the cumulative pieces of information he/she ever received. In this case, the
memory effect of non-Markovian process is induced.
On the other hand, the heterogeneity of effective spreading probability for edges
increases with the heterogeneity of degree distribution,
and further enhances the difficulty in developing an accurate theory.
The effective spreading probability of an edge includes two aspects: ($1$)
an edge is randomly selected with probability $f(k^\prime)/k^\prime$, where $k^\prime$ is the
degree of adopted individual $v$; ($2$) the information is transmitted through the
selected edge with probability $\lambda$. Thus, the effective spreading probability
of an edge for individual $v$ is $\lambda f(k^\prime)/k^\prime$.

To describe this process, we
develop a novel theory -- heterogeneous edge-based compartmental
theory, which is inspired by Refs.~\cite{Wang2014,Miller2011,Miller2013}.
The theory is based on the assumption that
behavior spreads on uncorrelated, large sparse networks. We denote $S(t)$, $A(t)$ and $R(t)$
as the density of individuals in the susceptible, adopted and recovered states at time $t$,
respectively.

Denoting $\theta_{k^\prime}(t)$ as the probability that an individual $v$ with
degree $k^\prime$ has not transmitted the information to individual $u$ along
a randomly selected edge up to time $t$. For simplicity, we assume that individuals
with identical degrees are the same in statistics. In the spirit of the cavity theory,
we let individual $u$ in the cavity state (i.e., individual $u$ can not transmit information
to his/her neighbors but can receive information from his/her neighbors).
Considering all possible degrees of individual $v$, the average probability
that individual $u$ has not received the information from his/her neighbors by time $t$
\begin{equation} \label{theta_mean}
\theta(t)=\sum_{k^\prime=0}\frac{k^\prime P(k^\prime)}{\langle k\rangle}\theta_{k^\prime}(t),
\end{equation}
where $k^\prime P(k^\prime)/\langle k\rangle$ represents the probability
that an edge from $u$ connects to
$v$ with degree $k^\prime$ in uncorrelated network, and $\langle k\rangle$ is the mean degree.
It is straightforward to get the probability that individual $u$ with $k$ neighbors has $m$ cumulate pieces of information by time $t$
\begin{equation} \label{phi_k_t}
\phi(k,m,t)=(1-\rho_0)\binom{k}{m}[\theta(t)]^{k-m}[1-\theta(t)]^m.
\end{equation}
The formula $1-\rho_0$ represents that only individuals in the susceptible state initially
can get the information. From Sec.~\ref{sec:model}, we know that only
when $u$'s cumulate pieces of information are less than $\kappa$, he/she can be
susceptible at time $t$. Thus, individual $u$ is susceptible by time $t$ with probability
\begin{equation} \label{S_K_T}
s(k,t)=\sum_{m=0}^{\kappa-1}\phi(k,m,t).
\end{equation}
Taking all possible values of $k$ into consideration, we can get the fraction (density) of
susceptible individuals at time $t$
\begin{equation} \label{S_T}
S(t)=\sum_{k}P(k)s(k,t).
\end{equation}
Similarly, we can get the fraction of individuals
who have received $m$ pieces of information at time $t$
\begin{equation} \label{sub_ind_S}
\Phi(m,t)=\sum_{k=0}P(k)\phi(k,m,t).
\end{equation}

According to the definition of $\theta_{k^\prime}(t)$, one can further divide it as
\begin{equation} \label{theta_k}
\theta_{k^\prime}(t)=\xi_{k^\prime}^S(t)+\xi_{k^\prime}^A(t)+\xi_{k^\prime}^R(t).
\end{equation}
The value of $\xi_{k^\prime}^S(t)$, $\xi_{k^\prime}^A(t)$, and $\xi_{k^\prime}^R(t)$
represents that the probability of individual $v$ with degree $k^\prime$ is susceptible,
adopted, and recovered and has not transmitted information to its
neighbors (e.g., individual $u$), respectively.

An initial susceptible neighbor individual $v$ of $u$ can only get the information from the
other $k^\prime-1$ neighbors, since individual $u$ is in the cavity state. Similar to
Eq.~(\ref{phi_k_t}), one can get the probability that $v$ has $m$ cumulate pieces of
information by time $t$
\begin{equation} \label{tao_S}
\tau(k^\prime,m,t)=(1-\rho_0)\binom{k^\prime-1}{m}[\theta(t)]^{k^\prime-m-1}[1-\theta(t)]^m.
\end{equation}
We further get the probability of individual $v$ in the susceptible
\begin{equation} \label{neighbour_S}
\xi_{k^\prime}^S(t)=\sum_{m=0}^{\kappa-1}\tau(k^\prime,m,t).
\end{equation}

If the adopted neighbor individual $v$ with degree $k^\prime$
transmits the information via an
edge, this edge will not meet the definition of $\theta_{k^\prime}(t)$. The conditions
of individual
$v$ transmits information to $u$ are: ($1$) the edge connecting them is selected with
probability $f(k^\prime)/k^\prime$ and ($2$) the information is transmitted through this
edge with probability $\lambda$. Thus, the evolution of $\theta_{k^\prime}(t)$ is
\begin{equation} \label{d_theta_k}
\frac{d\theta_{k^\prime}(t)}{dt}=-\frac{\lambda f(k^\prime)}{k^\prime}\xi_{k^\prime}^A(t).
\end{equation}
If $f(k^\prime)$ is larger than $k^\prime$, we restrict that $v$ transmits the information to his/her
all neighbors [i.e., $f(k^\prime)=k^\prime$].

According to information spreading process described in Sec.~\ref{sec:model}, the growth of
$\xi_{k^\prime}^R$ should simultaneously satisfy: ($1$) the adopted individual $v$ does
not transmit the information to $u$ through the edge between them and ($2$)
$v$ moves into recovered state with probability $\gamma$. For the first condition, there
are two possible cases: the edge between $u$ and $v$ is selected with probability
$f(k^\prime)/k^\prime$ and the information is not transmitted through it with probability
$1-\lambda$; the edge between $u$ and $v$ is not selected with probability
$1-f(k^\prime)/k^\prime$. From the analyses above, the evolution of
$\xi_{k^\prime}^R$ is
\begin{equation} \label{d_xi_R}
\frac{d\xi_{k^\prime}^R(t)}{dt}=\gamma\xi_{k^\prime}^A(t)[1-\frac{\lambda f(k^\prime)}{k^\prime}].
\end{equation}
Now, combining Eqs.~(\ref{d_theta_k})-(\ref{d_xi_R}) and the initial situations [i.e., $\theta_{k^\prime}(0)=1$
and $\xi_{k^\prime}^R(0)=0$], we obtain the expression of $\xi_{k^\prime}^R(t)$ in terms of $\theta_{k^\prime}(t)$ as
\begin{equation} \label{xi_R}
\xi_{k^\prime}^R(t)=\gamma[1-\theta_{k^\prime}(t)][\frac{k^\prime}{\lambda f(k^\prime)}-1].
\end{equation}
Utilizing Eqs.~(\ref{theta_k}), (\ref{neighbour_S}), (\ref{d_theta_k}) and (\ref{xi_R}),
we obtain that
\begin{equation} \label{d_theta_k_con}
\begin{split}
\frac{d\theta_{k^\prime}(t)}{dt}&=-\frac{\lambda f(k^\prime)}{k^\prime}[\theta_{k^\prime}(t)-\sum_{m=0}^{\kappa-1}\tau(k^\prime,m,t)]\\
&+\gamma[1-\theta_{k^\prime}(t)][1-\frac{\lambda f(k^\prime)}{k^\prime}].
\end{split}
\end{equation}

According to the model described in
Sec.~\ref{sec:model}, the densities of individuals in
adopted and recovered individuals evolve as
\begin{equation} \label{d_A_t}
\frac{dA(t)}{dt}=-\frac{dS(t)}{dt}-\gamma A(t)
\end{equation}
and
\begin{equation} \label{d_R_t}
\frac{dR(t)}{dt}=\gamma A(t),
\end{equation}
respectively. Eqs.~(\ref{S_T}) and (\ref{d_A_t})-(\ref{d_R_t}) give us a complete description of the
social contagions with limited contact capacity. The evolution of each type of
density versus time can be obtained.

The densities of susceptible, adopted and recovered individuals do not change when
$t\rightarrow\infty$. We denote $R(\infty)$ as the final behavior adoption size.
To obtain the value of $R(\infty)$, one can first solve $\theta_{k^\prime}(\infty)$ from
Eq.~(\ref{d_theta_k_con}), that is
\begin{equation} \label{d_theta_k_steady}
\begin{split}
\theta_{k^\prime}(\infty) =\sum_{m=0}^{\kappa-1}\tau(k^\prime,m,\infty)]
 +\gamma[1-\theta_{k^\prime}(\infty)][\frac{k^\prime}{\lambda f(k^\prime)}-1].
\end{split}
\end{equation}
Iterating Eq.~(\ref{d_theta_k_steady}) to obtain $\theta_{k^\prime}(\infty)$.
Then, inserting $\theta_{k^\prime}(\infty)$ into Eqs.~(\ref{theta_mean})-(\ref{S_T})
to get the values of $S(\infty)$ and $R(\infty)=1-S(\infty)$.

Another important aspect we mainly focus on is the condition under which the global behavior
adoption occurs. The global behavior adoption means that a finite fraction of individuals
adopted the behavior, and the corresponding local behavior adoption represents that
only a vanishingly small fraction of individuals adopted the behavior. Similar to biological
contagions, we define a critical transmission probability $\lambda_c$. When $\lambda\leq
\lambda_c$, the behavior can not be adopted by a finite fraction of individuals; when
$\lambda>\lambda_c$, the global behavior adoption occurs. Now, we discuss $\lambda_c$ for several different values of $\rho_0$ and $\kappa$.

\begin{figure}
\begin{center}
\epsfig{file=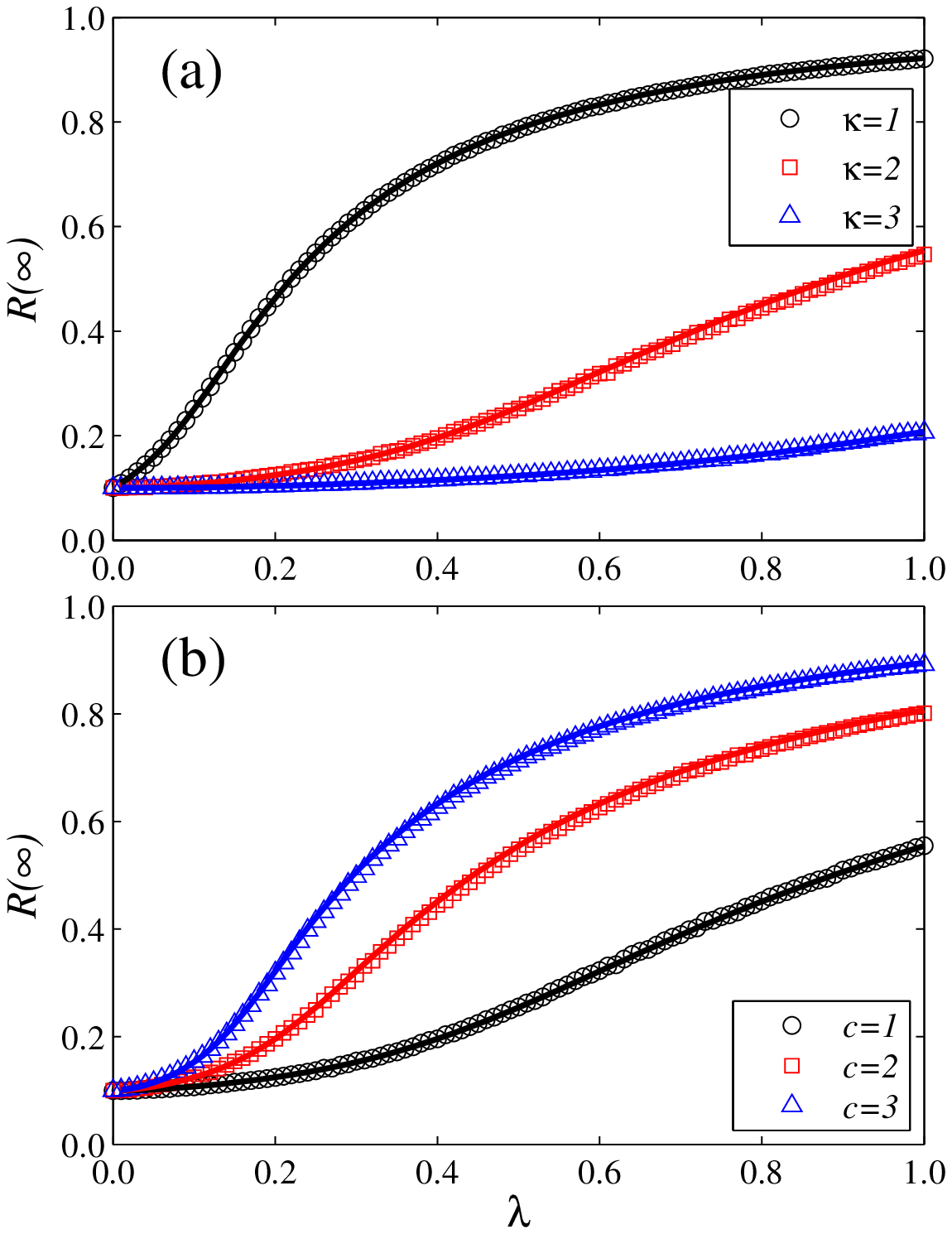,width=0.8\linewidth}
\caption{(Color online) On strong heterogeneous networks, the final adoption size
$R(\infty)$ as a function of information transmission probability $\lambda$ for
(a) different adoption threshold $\kappa$ and (b) different contact capacities
$c$. In (a), black circles ($\kappa=1$), red squares ($\kappa=2$) and blue up triangles
($\kappa=3$) are the simulation results for $c=1$. In (b), black circles ($c=1$),
red squares ($c=2$) and blue up triangles ($c=3$) are the simulation results for
$\kappa=2$. In figure (a) and (b), the lines are the theoretical predictions from
Eqs.~(\ref{S_T}) and (\ref{d_A_t})-(\ref{d_R_t}). We set other parameters as
$\nu=2.1$, $\gamma=0.1$, and $\rho_0=0.1$, respectively.}
\label{fig1}
\end{center}
\end{figure}

For $\rho_0\rightarrow0$ (i.e., only a vanishing small fraction of seeds)
and $\kappa=1$, $\theta_{k^\prime}(\infty)=1$ is the trivial solution of Eq.~(\ref{d_theta_k_steady}).
If we change the values of other dynamical parameters, such as information transmission
probability $\lambda$, a global behavior adoption may occur. The global behavior
adoption occurs only when a nontrivial solution of Eq.~(\ref{d_theta_k_steady}) emerges
[i.e., $\theta_{k^\prime}(\infty)<1$]. Note that the corresponding fraction of
$\theta_{k^\prime}(\infty)$ should be taken into consideration. Linearizing
Eq.~(\ref{d_theta_k_steady}) at $\theta_{k^\prime}(\infty)=1$~\cite{Newman2010}, and
summing all possible values of $k^\prime$ one can get the critical information transmission
probability
\begin{equation} \label{lambda_c}
\lambda_c=\frac{\gamma\langle k\rangle G(k)}{\langle k^2\rangle-(2-\gamma)\langle k\rangle},
\end{equation}
where
$$
G(k)=\sum_{k^\prime} \frac{k^{\prime^2} P(k^\prime)}{\langle k\rangle f(k^\prime)}.
$$
Note that $\lambda_c$ is tightly correlated with the network topology [i.e., degree distribution $P(k)$] and
dynamical parameters [i.e., contact capacity $f(k^\prime)$ and recover probability
$\gamma$]. Hubs in heterogeneous networks adopt the behavior with large probability.
Thus, the value of $\lambda_c$ decreases with degree heterogeneity.
The value of $\lambda_c$ increases with $f(k^\prime)$, in other words, increasing the contact capacity
of individuals makes the network more fragile to the behavior spreading. We should
emphasize that $\gamma$ also affects $\lambda_c/\gamma$ (i.e., the effective critical information transmission probability),
which has been neglected in previous studies. The value of $\lambda_c/\gamma$ increases with
$\gamma$. If $f(k^\prime)\geq k^\prime$ for every value of
$k^\prime$ (i.e., adopted individual transmits the information to his/her all neighbors),
Eq.~(\ref{lambda_c}) is the epidemic outbreak
threshold~\cite{Wang2014,Newman2002}. If every adopted
individual only transmits information to his/her $c$ neighbors, we can get the
critical transmission probability $\lambda_c$ of the model in Ref.~\cite{Yang2007}.

For $\rho_0\rightarrow0$ and $\kappa>1$, we find that $\theta_{k^\prime}(\infty)=1$ is
the solution of Eq.~(\ref{d_theta_k_steady}). However, the left and right hands of Eq.~(\ref{d_theta_k_steady}) can not be tangent to each other at $\theta_{k^\prime}=1$,
which indicates that vanishingly small seeds can not trigger the global behavior adoption~\cite{Wang2015a}.
With the increase of $\rho_0$, different dependence of $R(\infty)$ on $\lambda$ occurs for
different $\kappa$. That is, the growth pattern of $R(\infty)$ versus $\lambda$ can be
continuous or discontinuous. Through bifurcation analysis~\cite{Strogatz1994} of
Eq.~(\ref{d_theta_k_steady}), we find that $R(\infty)$ grows continuously for $\kappa=1$,
while a discontinuous growth may be induced for $\kappa>1$.

\section{Simulation Results}\label{sec:sim}
In this section, we verify the effectiveness of the heterogeneous edge-based compartmental theory developed in Sec.~\ref{sec:theory} by lots of simulations. For
each network, we perform at least $2\times 10^3$ times for a dynamic process and measure the
final fraction of individuals in the recovered [$R(\infty)$] and subcritical state [$\Phi(\kappa-1,\infty)$]. These results are then
averaged over 100 network realizations.

To built the network topology, we use the uncorrelated configuration model~\cite{Catanzaro2005}
according to the given degree distribution $P(k)\sim k^{-\nu}$ with maximal degree $k_{max}\sim
\sqrt{N}$. There is no degree-degree correlations when $N$ is very large. The heterogeneity of
network increases with the decrease of $\nu$. For the sake of investigating the effects of
heterogeneous structural properties on the social contagions directly, the network sizes and
mean degree are set to be $N=10,000$ and $\langle k\rangle=10$, respectively. All
individuals with different degrees have the same contact capacity $f(k)=c$ and recover
probability $\gamma=0.1$.

\begin{figure}
\begin{center}
\epsfig{file=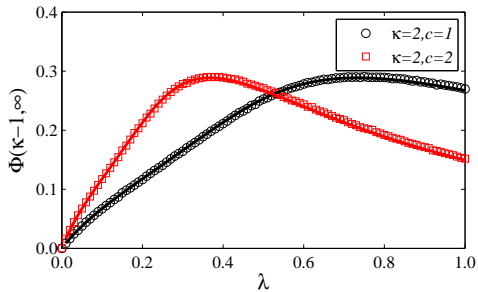,width=0.8\linewidth}
\caption{(Color online) The final fraction of individuals in the subcritical state $\Phi(\kappa-1,\infty)$
versus information transmission probability $\lambda$ for $\kappa=2$, $c=1$ (black circles) and
$\kappa=2$, $c=2$ (red squares). The lines are the theoretical predictions from Eqs.~(\ref{sub_ind_S}) and (\ref{d_A_t})-(\ref{d_R_t}).
Other parameters set to be $\nu=2.1$, $\gamma=0.1$, and $\rho_0=0.1$, respectively.}
\label{fig2}
\end{center}
\end{figure}

We first study the effects of the adoption threshold $\kappa$ and contact capacity $c$ on
the final behavior adoption size $R(\infty)$ for strong heterogeneous networks in
Fig.~\ref{fig1}.  We find that $R(\infty)$ decreases with the increase of $\kappa$,
since individuals adopting the behavior need to expose more information. Once the
contact capacity increases (i.e., $c$ increases), individuals in adopted state
will have more chances to transmit the information to susceptible individuals,
thus, the values of $R(\infty)$ increases. Obviously, the theoretical predictions
from heterogeneous edge-based compartmental theory agree well with the
simulation results.

\begin{figure}
\begin{center}
\epsfig{file=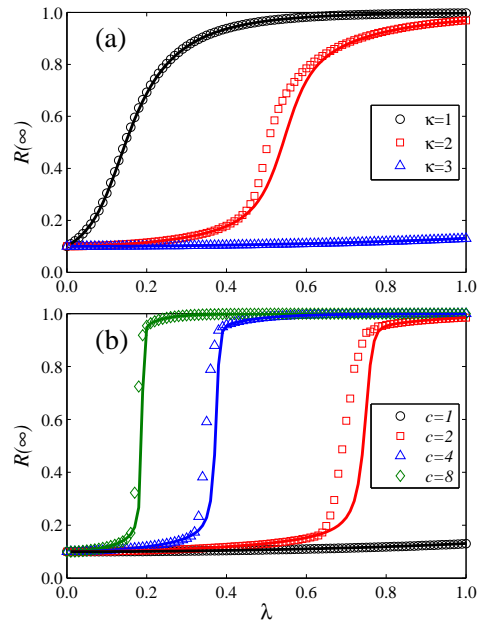,width=0.8\linewidth}
\caption{(Color online) On weak heterogeneous networks, the final adoption size
$R(\infty)$ versus information transmission probability $\lambda$ for (a) different
adoption threshold $\kappa$ and (b) different contact capacities
$c$. In (a), black circles ($\kappa=1$), red squares ($\kappa=2$) and blue up
triangles ($\kappa=3$) are the simulation results for $c=1$. In (b), black circles
($c=1$), red squares ($c=2$), blue up triangles ($c=4$) and green diamond ($c=8$)
are the simulation results for $\kappa=3$. And the lines are the theoretical
predictions, which are solved from Eqs.~(\ref{S_T}) and (\ref{d_A_t})-(\ref{d_R_t}).
We set other parameters as $\nu=4.0$, $\gamma=0.1$, and $\rho_0=0.1$.}
\label{fig3}
\end{center}
\end{figure}

Another important issue we concern is the dependence of $R(\infty)$ on $\lambda$.
As shown in Fig.~\ref{fig1}, for strong heterogeneous
networks the dependence of $R(\infty)$ on $\lambda$ is continuous for any
values of $\kappa$ and $c$, and we verify this claim by the bifurcation analysis of Eq.~(\ref{d_theta_k_steady}). We can also understand this
phenomenon by discussing the fraction of individuals
in the subcritical state from an intuitive perspective (see Fig.~\ref{fig2}).  An
individual in the subcritical state means that he/she is in the susceptible state,
and the $m$ cumulative pieces of
information is just one smaller than his/her adoption threshold $\kappa$.
From Ref.~\cite{Wang2015a}, we know that a discontinuous dependence of $R(\infty)$
on $\lambda$ will occur only when a large number of those subcritical
individuals adopt the behavior simultaneously at some information transmission probability.
Fig.~\ref{fig2} shows the final fraction of
individuals in the subcritical state $\Phi(\kappa-1,\infty)$ versus $\lambda$. We find that
$\Phi(\kappa-1,\infty)$ first increases and then decreases gradually with
$\lambda$, since the existence of strong degree
heterogeneity makes individuals in the subcritical state adopt the behavior consecutively.
In these cases, a continuous growth of $R(\infty)$ versus $\lambda$ occurs
on strong heterogeneous networks.

\begin{figure}
\begin{center}
\epsfig{file=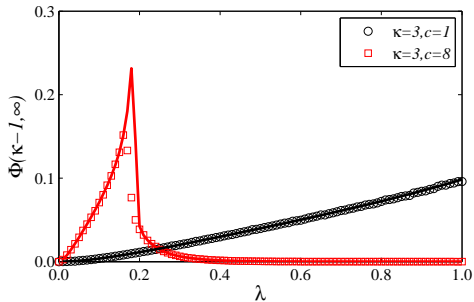,width=0.8\linewidth}
\caption{(Color online) The final fraction of individuals in the subcritical state $\Phi(\kappa-1,\infty)$
versus information transmission probability $\lambda$ for $\kappa=3$, $c=1$ (black circles) and
$\kappa=3$, $c=8$ (red squares). The lines are the theoretical predictions from Eqs.~(\ref{sub_ind_S}) and (\ref{d_A_t})-(\ref{d_R_t}).
Other parameters are $\nu=4.0$, $\gamma=0.1$, and $\rho_0=0.1$, respectively.}
\label{fig4}
\end{center}
\end{figure}

\begin{figure}
\begin{center}
\epsfig{file=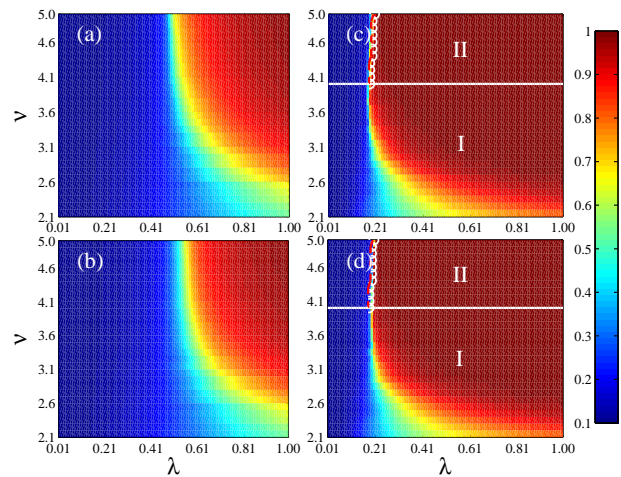,width=1\linewidth}
\caption{(Color online) The final behavior adoption size versus information transmission probability and degree exponent. (a) and (c) represent, respectively, the color-coded
values of $R(\infty)$ from numerical simulations in the parameter plane $\lambda$-$\nu$
for $c=1$, $\kappa=2$ and $c=8$, $\kappa=3$.  The theoretical predictions for
$c=1$, $\kappa=2$ and $c=8$, $\kappa=3$ are shown in (b) and (d), respectively.
And theoretical predictions are solved from Eqs.~(\ref{S_T}) and (\ref{d_A_t})-
(\ref{d_R_t}). In (b) and (d), $R(\infty)$ grows continuously with $\lambda$ to a
large value in region I. And in region II, $R(\infty)$ grows discontinuously and
a finite fraction of individuals adopt the behavior above the discontinuous information
transmission probability $\lambda_c^I$.
The horizontal white line is the critical degree exponent $\nu_c$, white circles and
red dashed lines
are simulated and theoretical results of $\lambda_c^I$, respectively.
Other parameters are $\gamma=0.1$ and $\rho_0=0.1$.}
\label{fig5}
\end{center}
\end{figure}

We now study behavior spreading on weak heterogeneous networks, such as $\nu=4.0$ in
Fig.~\ref{fig3}. Similar with the case of $\nu=2.1$, increasing $\kappa$ leads to the
decrease of $R(\infty)$; and the value of $R(\infty)$ increases with $c$, that is the network
will become more fragile to the behavior spreading once the contact capacity increases.
Once again, our theory can predict the social dynamics very well.  For the dependence of
$R(\infty)$ on $\lambda$, a crossover phenomenon transition is observed. A crossover
phenomenon means that the dependence of $R(\infty)$ on $\lambda$ can change from
being continuous to being discontinuous. More
specifically, as shown in Fig.~\ref{fig3}(b), the dependence of $R(\infty)$ on
$\lambda$ is continuous for small values
of $c$ (e.g., $c=1$), while the dependence is discontinuous for larger $c$
(e.g., $c=8$). We justify
this claim by the bifurcation analysis of Eq.~(\ref{d_theta_k_steady}) from the theoretical
view, which is also verified through analyzing $\Phi(\kappa-1,\infty)$ from
an intuitive perspective in Fig.~\ref{fig4}.  For weak
heterogeneous networks, most individuals adopt the behavior with the same probability
since they have similar degrees. When $c=1$, $\Phi(\kappa-1,\infty)$ increases
continuously with $\lambda$, which leads to a continuous growth in the value of $R(\infty)$.
When $c=8$, $\Phi(\kappa-1,\infty)$ first increases with
$\lambda$, and reach a maximum at some values $\lambda_c$, and a slight increment of $\lambda$ induces a finite fraction of $\Phi(\kappa,\infty)$ to adopt the behavior simultaneously, which
leads to a discontinuous jump in the value of $R(\infty)$.

We further study the effects of $\nu$ and $\lambda$
in Fig.~\ref{fig5} for different values of $c$. For small contact capacity
[i.e., $c=1$ in Figs.~\ref{fig5}(a) and (b)], the dependence of $R(\infty)$ on
$\lambda$ is always continuous for any value of $\nu$. In
other words, this dependence is irrelevant to the network topology.
For large contact capacity [i.e., $c=8$ in Figs.~\ref{fig5}(c) and (d)],
there is a crossover phenomenon
in which the dependence of $R(\infty)$ on $\lambda$ can change from being
continuous to being discontinuous. More particularly,
there is a critical degree exponent $\nu_c$ below which the dependence is continuous
[see region I in Figs.~\ref{fig5}(c) and (d)],
while above $\nu_c$ the dependence is discontinuous [see region II in Figs.~\ref{fig5}(c)
and (d)]. The value of $\nu_c$ can be gotten by bifurcation analysis of
Eq.~(\ref{d_theta_k_steady}). In region II, we also find that the discontinuous
information transmission probability $\lambda_c^I$ increases with $\nu$, since the fraction
of hubs
decreases with $\nu$. The theoretical predictions of $\lambda_c^I$ can be gotten
 by bifurcation analysis of Eq.~(\ref{d_theta_k_steady}), and the simulation results of
$\lambda_c^I$ are predicted by NOI (number of iterations) method~\cite{Wang2015a}.
Regardless of network heterogeneity, our theoretical predictions about the behaviors of
$R(\infty)$ have a good agreement with numerical calculations. The average relative
error~\cite{Glesson2012} between the two predictions of $R(\infty)$ for all the values of
$\lambda$ and $\nu$ is less than $1.8\%$.

\begin{figure}
\begin{center}
\epsfig{file=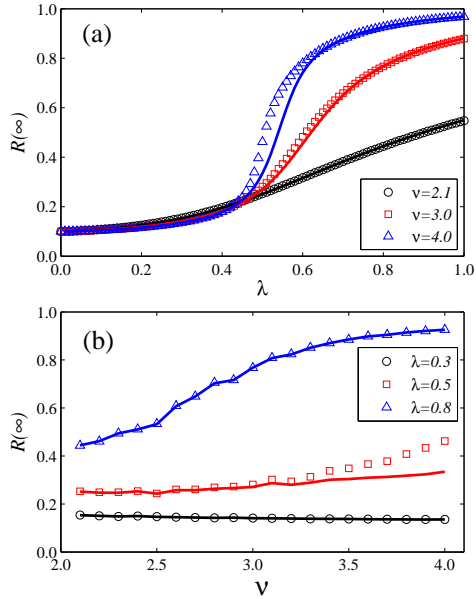,width=0.8\linewidth}
\caption{(Color online)
Behavior spreading on scale-free networks. (a)
The final behavior adoption size $R(\infty)$ versus information transmission probability
$\lambda$ for different degree exponents $\nu=2.1$ (black circles), $\nu=3.0$ (red squares) and
$\nu=4.0$ (blue up triangles).  (b) The final behavior adoption size $R(\infty)$ versus
$\lambda=0.3$ (black circles), $\lambda=0.5$ (red squares) and
$\lambda=0.8$ (blue up triangles), respectively. The lines are the theoretical predictions
from Eqs.~(\ref{S_T}) and (\ref{d_A_t})-(\ref{d_R_t}). Other parameters are $c=1$,
$\gamma=0.1$, $\kappa=2$, and $\rho_0=0.1$, respectively.}
\label{fig6}
\end{center}
\end{figure}

Finally, we study the effects of network topology on the final behavior adoption size
$R(\infty)$ in Fig.~\ref{fig6} for $\kappa=2$ and $c=1$. We find that increasing $\nu$
can promote (suppress) behavior adoption at large (small) value of $\lambda$. This
phenomenon can be qualitatively understood in the in following ways~\cite{Wang2014,Wang2015a}:
For strong heterogeneous networks, the more hubs and a large number of
individuals with small degrees are coexisted. Since those hubs have more chances to
expose the information, they adopt the behavior more easily even when $\lambda$ is
small. However, the situations for individuals with small degree are just opposite.
That is, the large number of individuals with small degrees hinder the behavior adoption
for large value of $\lambda$, thus, cause a smaller value
of $R(\infty)$. Through bifurcation analysis of Eq.~(\ref{d_theta_k_steady}), the
dependence of $R(\infty)$ on $\lambda$ is always continuous for different
$\nu$.

\section{Discussion} \label{sec:dis}
To study social contagion dynamics in human populations
is an extremely challenging problem with broad implications and interest. For social contagions on networks, some inelastic resources (e.g., time, funds, and
energy) restrict individuals to dedicate to social interaction, which have always been
neglected in previous studies. In this paper, we first proposed a non-Markovian behavior
spreading model with limited contact capacity, in which each adopted individual transmits
the information to a fraction of his/her neighbors. We then developed a heterogeneous
edge-based compartmental theory to describe this model. The
average relative error between the theoretical
predictions and numerical calculations is less than $1.8\%$. Through theory and simulations,
we found that increasing the contact capacity $c$ promotes the final behavior adoption
size $R(\infty)$. With the help of bifurcation theory, we
found a crossover phenomenon in which the dependence
of $R(\infty)$ on $\lambda$ can change from being continuous to being
discontinuous. More specifically, we uncovered a critical degree exponent $\nu_c$ above which the crossover phenomenon can be induced by enlarging $c$. However,
$R(\infty)$ always grows continuously for any value of $c$ when degree exponent is below
$\nu_c$.

Here we developed an accurate theoretical framework for non-Markovian social contagion model
with limited contact capacity, which could be applied to other analogous dynamical processes (e.g.,
information diffusion and cascading failure). Further more, how to design an effective
strategy to control the behavior spreading with limited contact capacity is an
interesting research topic.

\acknowledgments

This work was partially supported by the National Natural Science
Foundation of China under Grant Nos.~11105025, 91324002 and
the Program of Outstanding Ph.~D. Candidate in Academic Research by
UESTC under Grand No.~YXBSZC20131065.

%\bibliographystyle{apsrev4-1}
%\bibliography{Social_Contagion}

\end{document}